\documentclass[letter]{aa}
\usepackage{graphicx}
\usepackage{natbib,txfonts}                               
\usepackage{url}
\usepackage{soul}
\usepackage{changes}

\def\kms{km\,s$^{-1}${}}
\def\mkm{$\mu$m}
\nolinenumbers

\begin{document}

\title{The jet and circumstellar environment of the young binary DF~Tau }
\titlerunning{The jet of DF~Tau}
\authorrunning{Dodin et al.}

   \author{A.V.~Dodin, M.A.~Burlak, V.A.~Kiryukhina, S.A.~Lamzin, I.A.~Shaposhnikov, \\ I.A.~Strakhov, A.A.~Tatarnikov, A.M.~Tatarnikov
          }

   \institute{Sternberg Astronomical Institute, Lomonosov Moscow State University,
              Universitetskij prospekt 13, 119234 Moscow, Russia\\
              \email{lamzin@sai.msu.ru}
             }

   \date{Received ...; accepted ...}

  \abstract
   {Jets and disc winds play an important role in the 
   evolution of protoplanetary discs and the formation 
   of planetary systems. However, there is still a lack of observational data regarding the presence and parameters of outflows, especially for close young binaries.}
   {In this study, we aim to find the HH flow near the young sub-arcsecond binary 
   DF~Tau and explore its morphology.}
   {Narrow-band H$\alpha$ and H$_2$~2.12~\mkm{} imaging and 
   spectroscopic observations of DF~Tau and its vicinity were 
   performed. }
   {We have discovered several emission nebulae near the binary, which likely result from the interaction of gas outflow from the binary components with the surrounding medium. The outflow appears to occur both in the form of jets, generating numerous Herbig-Haro objects (HH 1266 flow), and as a weakly collimated wind responsible for the formation of the ring-like nebula around the binary and the rim of the cometary globule. We have found that the angle between the jet and the counter-jet is $168\degr$ and discuss the complex morphology of the HH flow.}
  {}


   \maketitle

\section{Introduction}
 \label{sec:intro}

Classical T~Tauri stars (CTTSs) are young (age $< 10^7$~yr), low mass ($M \lesssim 2.5$~M$_\odot)$ stars at the stage of gravitation contraction towards the main sequence, whose activity is driven by the magnetospheric accretion of matter from the protoplanetary disc \citep{BBB-1988,Hartmann-2016}. The accretion process is accompanied by the outflow of disc matter in the form of a weakly collimated disc wind and, in some cases, jets. The outflowing matter reduces the mass of the disc and carries away its angular momentum, thereby setting the initial conditions for planet formation \citep{Pascucci-PPVII-2023}. The interaction of the disc wind and jet with the remnants of the protostellar cloud largely determines the morphology of environment around young stars \citep{Frank-2014}. 

Jets are extended (up to 3 pc), highly supersonic $(V \sim 300$~\kms), collimated bipolar gas flows \citep{Bally-2016}. They appear as a chain of compact emission nebulae (knots), known as Herbig-Haro (HH) objects, with less dense gas between them. There is a general consensus that the large scale magnetic field of CTTSs and its interaction with the protoplanetary disc are responsible for the jet collimation and acceleration -- see, for example, \citet{Sheikhnezami-2024} and references therein. Nevertheless, a number of issues related to the details of jet formation remain poorly known, particularly the size of the disc area from which the jet is launched \citep{Ferreira-2006}. 

In this regard, it is important to search for and explore jets of CTTSs in close binary systems (semimajor axis $a\lesssim 20$~au), since the tidal interaction truncates the radius of the companion's circumstellar disc to $R_{\text out} \sim 0.4\,a$ \citep{Artymowicz-Lubow-1994, Rosotti-Clarke-2018}.

We report here the discovery of an extended, non-trivially shaped jet from the sub-arcsecond young binary DF~Tau. The star attracted attention after \citet{Joy-1949} discovered emission lines in its spectrum and identified it as a M0 T~Tauri type star. Soon afterwards, \citet{Kholopov-Kurochkin-1951} found that DF~Tau is a strongly variable star $(\Delta m_{\text pg}> 2^{\text m})$. According to \citet{Bailer-Jones-2021}, the Gaia parallax for DF~Tau (DR3~151179966897747840) has a large error, with RUWE$\approx 22$ \citep{Gaia-collaboration-2021}. Therefore, we will further use the distance $D=140$~pc as the average distance to the D4-North subgroup where DF~Tau resides \citep{Krolikowski-2021, Kutra-2025}.

\citet{Chen-1990} found DF~Tau to be a binary system consisting of two CTTSs \citep{White-Ghez-2001} of approximately equal mass and M2.5 spectral type \citep{Hartigan-Kenyon-2003}. According to \citet{Kutra-2025}, the effective temperatures of DF~Tau~A and B components are $3640 \pm 100$\,K and $3430 \pm 80$\,K, respectively, and both have a global magnetic field of $\approx 2.5$~kG. The orbital period, semimajor axis and eccentricity of the system are approximately 50~yr, $0\farcs 1,$ and $0.2$, respectively \citep{Allen-2017, Kutra-2025}, which corresponds to a minimum distance between the components of $< 12$~au. 

DF~Tau is the source of an outflow observed in forbidden \citep{HEG-1995, Nisini-2023} and dipole-allowed \citep{Edwards-1994, Lamzin-HST-2001} transitions. \citet{Hartigan-Jet-2004} and \citet{Uvarova-Jet-2020} concluded that micro-jets with a projected length of $<0\farcs 2$ exist on both sides of the binary, {\lq}either as a jet and its counter-jet or as separate jets from the primary and secondary{\rq}. Considering all the above, the study of large-scale matter outflow in the vicinity of DF~Tau is reasonable.

\section{Observations}
 \label{sec:obs}

The observations of DF~Tau were carried out with the 2.5~m telescope of the Caucasian Mountain Observatory (CMO) of the Sternberg Astronomical Institute of the Lomonosov Moscow State University (SAI MSU) \citep{Shatsky-2020}. Spectroscopic data were obtained using the Transient Double-beam Spectrograph (TDS) -- see \citet{Potanin-2020} for a description of the instrument and data reduction procedures. The spectral resolving power of the TDS with the $1\arcsec$-slit is $R=\lambda / \Delta \lambda \approx 2400$ in the red channel $(0.56 - 0.74$~\mkm) and $\approx 1300$ in the blue one $(0.36 - 0.56$~\mkm). The log of observations is given in Table~\ref{tab:log-TDS}, where $\text{rJD}={\text JD}-2\,460\,000$, ${\text s}_{\text i}$ refers to the slice numbered $i$ in the right panel of Fig.~\ref{fig:vicinity}, and PA is the position angle. A slit of $3\arcmin$ length and $1\arcsec$ width was used in all cases, except the spectrum of DF~Tau itself, where it was $3\arcmin\times10\arcsec$.

\begin{table}
\caption{Log of TDS observations}
\label{tab:log-TDS}      
\centering                   
\begin{tabular}{l c c c}        
\hline\hline
ID &     rJD   & slit PA, $\degr$  &  Exposure, s \\
\hline
DF~Tau & 588.6 & 128 & 2490 \\
$\rm s_1$  & 651.6 & 0  & 4800    \\
$\rm s_2$  & 654.5 & 0  & 7200    \\
$\rm s_3$  & 675.4 & 45  & 7200     \\
$\rm s_4$  & 704.4 & -79  & 4800     \\
$\rm s_5$  & 707.3 & 45  &  7200    \\
$\rm s_6$  & 729.2 & 0  &  3600    \\
\hline
\end{tabular}
\end{table}

Direct $10\arcmin \times 10\arcmin$ images of the region around DF~Tau were obtained with the 4K$\times$4K CCD camera of the 2.5-m telescope in two filters --- Halp~656~nm $(\lambda_{\text c}=656$~nm, $W=7.7$~nm) with a total exposure of $\Delta t=80$~min and nearby continuum Halpbc $(\lambda_{\text c}=643$~nm, $W=12$~nm, $\Delta t= 40$~min) -- on Dec~3, 2024 (rJD=648.3) and Jan~27, 2025 (rJD=703.3). \footnote{\url{https://obs.sai.msu.ru/cmo/sai25/wfi/}} 

Using the same telescope and the infrared camera ASTRONIRCAM \citep{Nadjip-17}, we also obtained images of the same regions in January-February 2025 in H2 $(\lambda_{\text c}=2.129$~\mkm, $W=46$~nm), $\Delta t= 175$~min) and Kcont $(\lambda_{\text c}=2.270$~\mkm, $W=39$~nm), $\Delta t=108$~min) filters. The details of the observations and data reduction are described in \citet{Dodin-2019}.

\section{Results and discussion}
 \label{sec:results}

   \begin{figure*}
   \centering
   \includegraphics[width=\hsize]{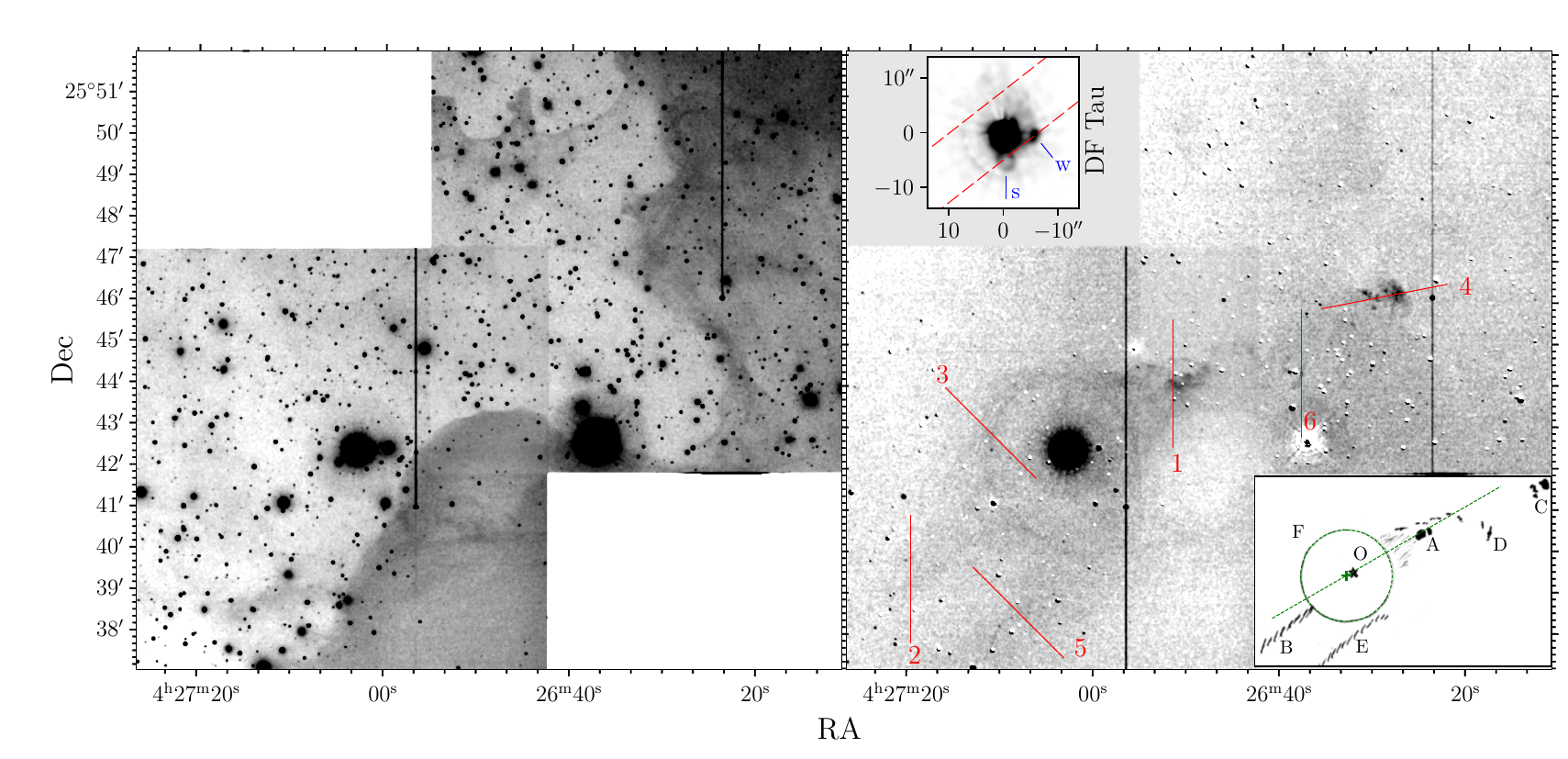}
      \caption{The vicinity of DF~Tau in the continuum band Halpbc (left panel), and the difference between images in the Halp 656~nm and Halpbc filters (right panel). Red segments indicate the position and orientation of the TDS slits. The close vicinity of DF~Tau is highlighted in a $10\arcsec \times 10\arcsec$ subimage located in the upper-left part of the panel. Red dashed lines mark the edges of the $10\arcsec$-slit used for TDS observation at $\rm rJD=588.6,$ southern (s) and western (w) blobs are marked with blue lines. The diagram in the lower-right part of the panel illustrates the nomenclature of nebulae designations used in the text: {\bf O} -- DF~Tau; green line -- micro-jet direction $({\text PA}=-59\degr$); {\bf A} -- the first group of bright HH objects; {\bf B} -- counter-jet, {\bf C} -- the second group of bright HH objects; {\bf D} -- 
      the brightest HH object ($\sim 3\times10^{-16}$ erg s$^{-1}$ \AA$^{-1}$ cm$^{-2}$ arcsec$^{-2}$); {\bf E} -- the rim; {\bf F} -- the ring, with its center marked by a green cross near DF~Tau.
      }
         \label{fig:vicinity}
   \end{figure*}

A mosaic of two images of DF~Tau's vicinity observed in the Halpbc filter (left panel) and the difference between the Halp and Halpbc filter images of the same region (right panel) are shown in Fig.~\ref{fig:vicinity}. DF~Tau resides in a (relatively) low CO and H$_2$ column density region between the B213 and B18 dark clouds \citep{Onishi-1996}. The south-eastern edge of the B213 cloud can be seen in the western part of the Halpbc image, as well as cometary globule 1 \citep[][Fig.17 and Table 6] {Goldsmith-2008} to the south-west of DF~Tau. A comparison of the images in the left and right panels of Fig.\ref{fig:vicinity} reveals that these areas are mostly reflection nebulae.

The right panel of the figure also allows us to identify a number of H$\alpha$-emitting nebulae in the vicinity of the binary. To understand their nature, we determined the {\it flux-averaged heliocentric} radial velocities $(V_{\text r})$ of the nebulae, using spectra observed at several characteristic points -- see Fig.\ref{fig:vicinity}, \ref{fig:pvd}, \ref{fig:spec1D} and Table~\ref{tab:log-TDS}. We will compare these velocities with the center-of-mass velocity of DF~Tau $(V^{\text cm}_{\text r}=+13\pm1$~\kms, \citealt{Allen-2017}). The diagram in the bottom-right corner of Fig.~\ref{fig:vicinity} (the scheme) illustrates the nomenclature of nebula designations used in this paper.

\subsection{The ring and the filament}
 \label{subsec:ring}

\begin{figure}
   \centering
   \includegraphics[width=\hsize]{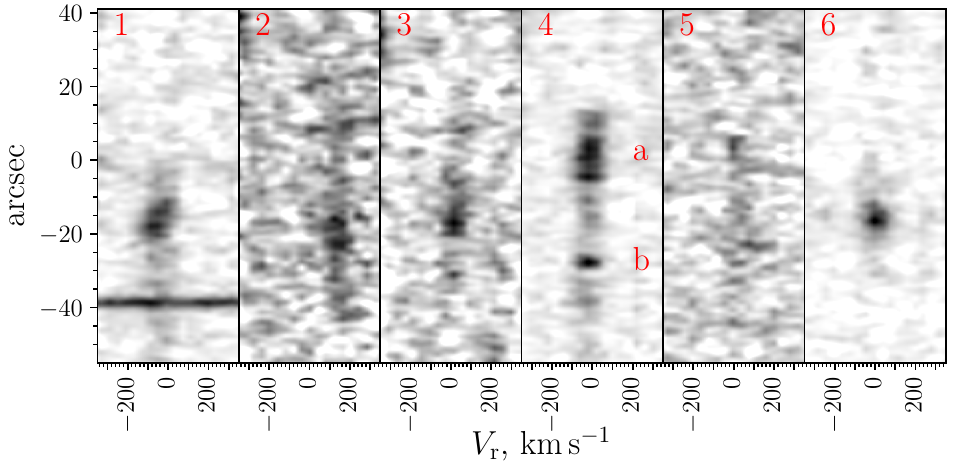}
      \caption{Position-velocity diagrams (PVDs) corresponding to slit positions marked in Fig.~\ref{fig:vicinity}. The horizontal strip in the left panel is the spectrum of a field star. Letters {\bf a} and {\bf b} correspond to different HH objects that fell within the spectrograph slit ($\rm s_4$ in Fig.\ref{fig:vicinity}).}
         \label{fig:pvd}
   \end{figure}

The first thing that catches the eye is the ring-shaped nebula around DF~Tau (letter {\bf F} in the scheme of Fig.\ref{fig:vicinity}). The radius of the ring is about $110\arcsec,$ which corresponds to a projected distances of $\approx 1.5\times10^4$~au. Judging by the TDS spectrum obtained at slit position 3 (see Fig.~\ref{fig:vicinity} and \ref{fig:pvd}), the gas in the ring moves with a radial velocity $V_{\text r}$ of $+24\pm 6$~\kms, indicating that it is moving away from the star at a speed of $\sim 10$~\kms.

The center of the ring is shifted by $\approx 10\arcsec$ to the south-east from the observed position of the binary (Fig.\ref{fig:vicinity}). The proper motion of DF~Tau is practically unknown due to the large errors in the Gaia astrometric solution, so we lack the information needed to interpret this fact.
 
H$\alpha$-emitting ring could be either an \ion{H}{ii} region or a shock wave resulting from the interaction of poorly collimated wind of DF~Tau's components with the background gas. The flux ratio $\xi$ of the hydrogen H$\alpha$ and H$\beta$ lines would help distinguish between these interpretations: given that the extinction to DF~Tau is $A_{\text V} \eqslantless 0.6$ \citep{Hartigan-Kenyon-2003, Herczeg-14}, a value of $\xi > 3$ would strongly support the shock wave interpretation \citep[][Fig.~18]{Sutherland-Dopita-2017}. Unfortunately, our $\rm s_3$ slice spectrum of the ring is very noisy, and the most we can conclude is that $\xi>2.$ 

We found that the H$\alpha$ flux from the ring, $F_\alpha$, is $\sim 2\times 10^{-11}$~erg~s$^{-1}$~cm$^{-2},$ so the ring's luminosity is $L_\alpha = 4\pi D^2 F_\alpha \sim 5\times10^{31}$~erg~s$^{-1}$. It is clear that components of the binary, with $T_{\text eff}<3\,800$~K and bolometric luminosity $L_{\text bol}<0.7$~L$_\odot$ \citep{Allen-2017} cannot produce enough Lyman continuum photons to sustain a {\it stationary} \ion{H}{ii} region with such a high H$\alpha$ luminosity. In principle, it is possible that the \ion{H}{ii} region arose in the past as a result of short-duration flare(s) of UV radiation, and the observed H$\alpha$ emission is the result of subsequent recombination.

Extreme flare-like brightenings of DF~Tau $(\Delta B > 4^{\text m})$ were observed twice in the last century \citep{Lamzin-2001}, apparently once per orbital period ($P\approx 50$~yr) of the binary. However, the spectra of the Year 2000-flare \citep{Li-2001} do not give reason to suspect that powerful Lyman continuum emission occurred during this event, which lasted about a day.

We therefore believe that the ring is not an \ion{H}{ii} region, but its H$\alpha$ emission results from the interaction of a poorly collimated gas outflow from DF~Tau’s components with the ambient matter. It is possible that the ring is actually a projection of a spherical bubble onto the celestial sphere. In this regard, it is worth noting that several dozen bubbles have been observed in the Taurus star-forming region in CO molecular lines, which are interpreted as {\lq}the manifestation of strong stellar winds dispersing the surrounding gas{\rq} \citep{Li-bubble-Taurus-2015}.

   \begin{figure}
   \centering
   \includegraphics[width=\hsize]{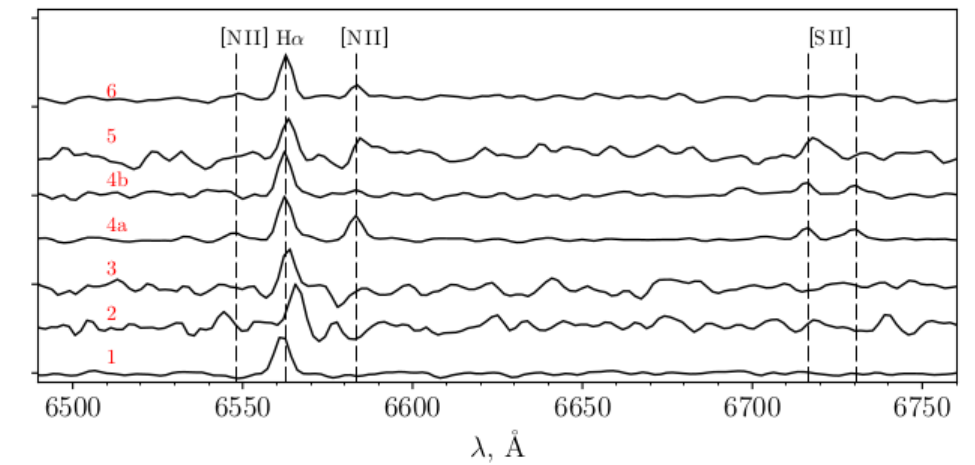}
      \caption{1D spectra corresponding to the PVDs in Fig.~\ref{fig:pvd}. The numbers 4a and 4b mark the spectra corresponding to the HH objects {\bf a} and {\bf b} in Fig.~\ref{fig:pvd}.}
         \label{fig:spec1D}
   \end{figure}

The lines of the [\ion{S}{II}]~6716+6731~\AA{} doublet are not visible in our spectrum of the ring ($\rm s_3$ in Fig.~\ref{fig:spec1D}), so we cannot estimate the electron number density $N_{\text e}$ or the ring's kinetic energy. Without this information, it is difficult to determine the relationship between the H$\alpha$-emitting ring of DF~Tau and the {\lq}molecular{\rq} bubbles found by \citet{Li-bubble-Taurus-2015}.

We believe that the filament (the rim) running along the north-eastern side of the cometary globule (letter {\bf E} in in the scheme of Fig.~\ref{fig:vicinity}) has the same origin as the ring, namely, it is the result of the interaction of DF~Tau's wind with the ambient medium. The H$\alpha$ flux of the rim is 3-4 times less than that of the ring. Judging by slice 5, the heliocentric $V_{\text r}$ of the rim is $+26\pm9$~\kms. As can be seen from the $\rm s_5$ spectrum in Fig.~\ref{fig:spec1D}, the spectrum of the rim is noisy, so it is unclear whether the [\ion{N}{ii}]~6583~\AA{} line is indeed present. However, we believe that the [\ion{S}{ii}]~6716~\AA{} line is observed and significantly stronger than the [\ion{S}{ii}]~6731~\AA{} line, which indicates \citep{Proxauf-SII-ratio-2014} that $N_{\text e} \lesssim 30$~cm$^{-3}$ in this part of the rim.

\subsection{The jet and HH~1266 flow}
 \label{subsec:jet}

\citet{Hartigan-Jet-2004} found that DF~Tau {\lq}shows clear jets at ${\text PA}\approx 127\degr$ and $\approx 307\degr${\rq}. More precisely, they identified elongated [\ion{O}{i}] emission features at a distance of $\approx 0\farcs 2$ from the binary (see also \citealt{Uvarova-Jet-2020}).

\begin{figure}
   \centering
   \includegraphics[width=\hsize]{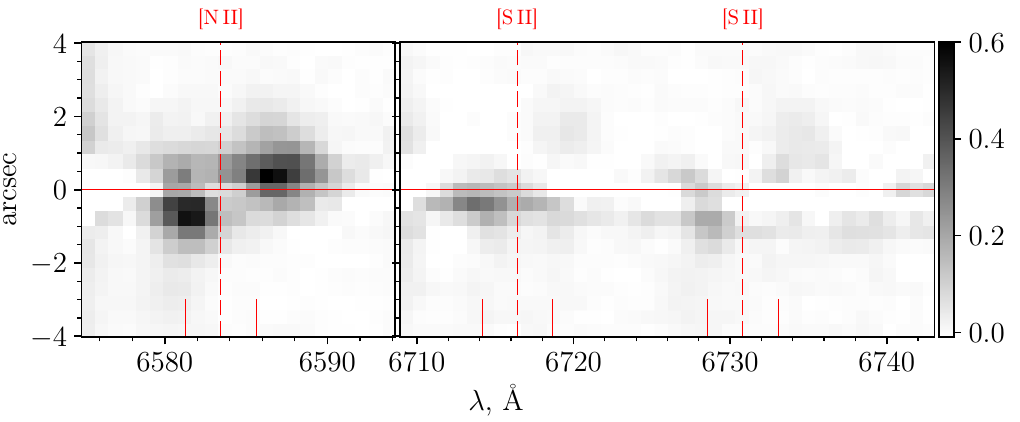}
      \caption{Long-slit spectra of DF~Tau with ${\rm PA}=128\degr.$ 
      The underlying stellar spectrum has been removed. Rest velocities and 
      $\pm 100$\,\kms relative to the stellar velocity are marked for 
      selected forbidden lines with red vertical lines. The colour bar 
      shows the flux in pixels ($0\farcs 366 \times 0\farcs 366$) relative to 
      the stellar spectrum, expressed as percentages.
              }
         \label{fig:micro}
   \end{figure}

A portion of our long-slit spectrum of DF~Tau, observed on October 5, 2024 (see table~\ref{tab:log-TDS}), is shown in Fig.~\ref{fig:micro}. Emission features are clearly visible in the [\ion{N}{ii}]~6583~\AA{} line, located on opposite sides of the star at a distance of $\approx 0\farcs 5$ and moving in opposite directions with a radial velocity $V_{\text r}$ of $\approx 100$~\kms. This is likely the micro-jet discovered by \citet{Hartigan-Jet-2004} in the [\ion{O}{i}]~6300 and 6363~\AA{} lines. However, we were unable to reliably separate the micro-jet's emission from the telluric oxygen lines in our spectrum, and therefore cannot estimate the proper motion of the micro-jet, as the line-forming region may differ for the [\ion{O}{i}] and [\ion{N}{ii}] lines \citep{Nisini-2023}. 

The [\ion{S}{ii}]~6716 and 6731~\AA{} lines are also present in our spectrum, but only in the part of the micro-jet approaching us. The radial velocity $V_{\text r}$ of the [\ion{S}{ii}] lines is $\approx -100$~\kms. The intensity of the red component of the sulfur doublet is noticeably less than that of the blue one, which indicates \citep{Proxauf-SII-ratio-2014} that $N_{\text e}\lesssim300$~cm$^{-3}$ in the line-forming region. 
 
Within the measurement errors, the direction of the micro-jet is perpendicular to the node line of DF~Tau's orbit \citep{Allen-2017, Kutra-2025}, as expected if the micro-jet is directed perpendicular to the orbital plane. The direction of the micro-jet is shown by the green line {\bf OA} in the scheme of Fig.~\ref{fig:vicinity}. The line with PA$_{\text j}=301\degr$ passes through group {\bf A} of emission nebulae. The PVD and spectrum of the brightest of these nebulae (slice $\rm s_1$) are shown in Fig.~\ref{fig:pvd} and \ref{fig:spec1D}, respectively. The $V_{\text r}$ of this nebulosity is $-56\pm 5$~\kms, which means that the nebula is moving away from the binary with $V_{\text r}\approx 70$~\kms. 

We conclude therefore that this part of the outflow is a jet from DF~Tau, and the group {\bf A} H$\alpha$-emitting nebulae are HH objects connected with the jet, as well as other nebulae located to the north-west of point {\bf A}, as we will see below. Prof. Bo~Reipurth agreed with our arguments and included the HH flow from DF~Tau, named as HH~1266, in his catalog \citet{Reipurth-2000}.

The HH~1266 flow has two remarkable features. Firstly, we found that the radial velocity $V_{\text r}$ of the elongated nebulosity {\bf B} (see the scheme in Fig.~\ref{fig:vicinity}) is $+138\pm 6$~\kms (Fig.\ref{fig:pvd} and \ref{fig:spec1D}), indicating that it moves away from the binary with a $V_{\text r}$ of $\approx 125$~\kms, and thus can be interpreted as a counter-jet with a projected length of $5\farcm 5$. Its $V_{\text r}$ is approximately twice as large as that of the jet. However, such velocity asymmetry between jets and counter-jets is observed in about half of all bipolar jets and is therefore very common \citep{Hirth-1994}. More intriguing is the fact that the counter-jet direction $({\text PA}\approx133\degr)$ differs by $12\degr$ from PA$_{\text j}-180\degr$. Moreover, it is clear even upon visual inspection in Fig\ref{fig:vicinity} that the jet and counter-jet are not aligned in diametrically opposite directions.

This is possible if the jet and counter-jet are launched by different components of the DF~Tau binary. According to \citet{Kutra-2025}, the inclinations of the circumprimary and circumsecondary discs of DF~Tau relative to the orbital plane are $13\degr \pm 13\degr$ and $8\degr \pm 9\degr,$ respectively. The authors concluded that {\lq}the disc-orbit obliquity is consistent with zero for both components{\rq}, but this does not exclude the possibility of the internal regions of the components discs being tilted relative to each other at an angle of $5-10\degr.$

   \begin{figure}
   \centering
   \includegraphics[width=\hsize]{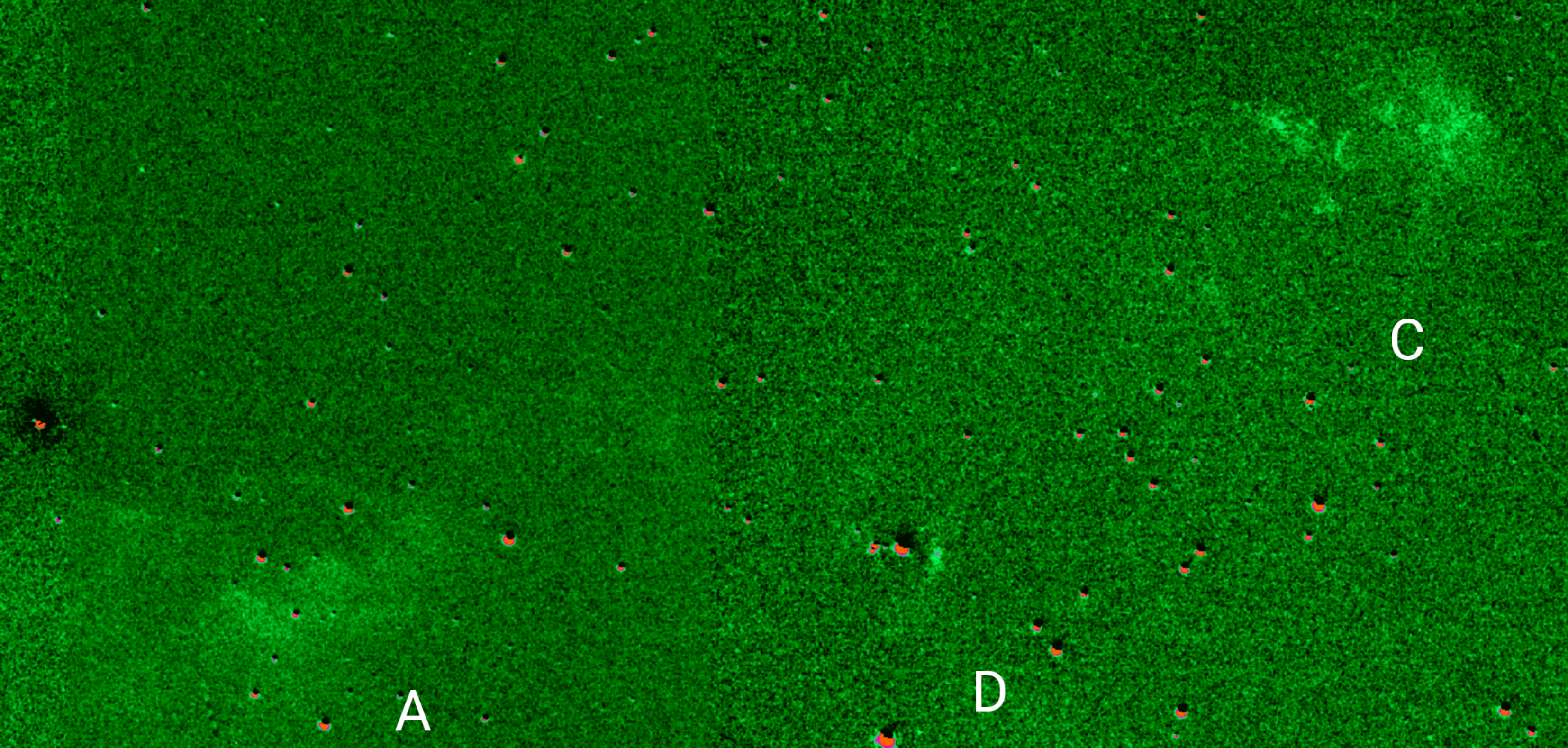}
      \caption{Large-scale vicinity of the jet. The letter designations are the same as in the scheme in Fig.\ref{fig:vicinity}.}
         \label{fig:jet-colour}
   \end{figure}

One more interesting feature of the HH~1266 flow is the relative location of the emission nebulae. Fig.\ref{fig:jet-colour} shows the north-western part of the jet in an enlarged view, with letter designations the same as in the scheme in Fig.\ref{fig:vicinity}. It can be seen that the emission nebulae are not arranged along a single smooth curve. This may be due to the fact that DF~Tau's components were launching jets in different directions, or/and jet deflection occurred due to its interaction with a non-homogeneous ambient medium -- see, for example, \citet{Raga-jet-deflection-2025} and references therein. Information about the proper motion of HH objects is necessary to interpret the observed morphology of the HH~1266 flow. We note that there are no young stellar objects within $0.5\degr$ of DF~Tau \citep{Luhman-TTau-list-2018}.

Our spectra of two {\bf C} group nebulae, located $\approx 10\farcm 5$ away from the star, which corresponds to a projected distance of $r_{\text C} \approx 10^5$~au (Fig.\ref{fig:vicinity}, \ref{fig:jet-colour} and the $\rm s_{4a}$, $\rm s_{4b}$ spectra in Fig.\ref{fig:pvd}, \ref{fig:spec1D}), indicate that they are indeed HH objects, moving with approximately equal radial velocities $V_{\text r}$ of $\approx -22\pm4$~\kms. Judging by the flux ratio of the [\ion{S}{ii}] lines, $N_{\text e}\sim 300$~cm$^{-3}$ in both objects. The $[\ion{N}{ii}]$~658~nm line is very weak if present at all in the 4b spectrum but has nearly the same intensity as the sum of [\ion{S}{ii}] doublet lines in the spectrum of the 4a HH object. If we also take into account that the flux ratio of H$\alpha$ and [\ion{O}{i}]~630~nm lines in the $\rm s_{4a}$ spectrum is $1.84 \pm 0.35$, then we find that the velocity of the shock exciting this HH object is $V_{\text sh}>30$~\kms \citep[][Fig.6]{Dopita-2017}. Therefore, an {\it order-of-magnitude estimation} of the {\bf C} group age is $r_{\text C}/V_{\text sh} \lesssim 2\times 10^4$~yr.

The spectrum of the {\bf D} nebula ($\rm s_6$ in Fig.\ref{fig:spec1D}) resembles that of the 4a HH object, but the flux ratio of the $[\ion{N}{ii}]$~658~nm line to the [\ion{S}{ii}] doublet lines is $>3$, implying that the excitation shock velocity is $>50$~\kms. Unfortunately, we were unable to separate the [\ion{O}{i}]~630~nm line of this HH object from the telluric line to confirm this conclusion. The radial velocity $V_{\text r}$ of this HH object is $-3 \pm 4$~\kms, which is close to that of the background CO molecular gas: $+6\pm 2$~\kms \citep{Li-bubble-Taurus-2015}. We believe therefore that the HH object is a terminated shock resulting from the collision of a low-density jet with a much denser ambient cloud \citep{Raga-book-2020}.

In conclusion of this section, we note that we attempted to detect H$_2$ 2.12~\mkm{} line emission from the HH~1266 flow region (see Sect.\ref{sec:obs}); however, only a $1\,\sigma$ upper limit of $7.5\times10^{-17}$~erg\,s$^{-1}$\,cm$^{-2}$\,arcsec$^{-2}$ was obtained.

\subsection{The blob}
 \label{subsec:bubble}

\begin{figure}
   \centering
   \includegraphics[width=\hsize]{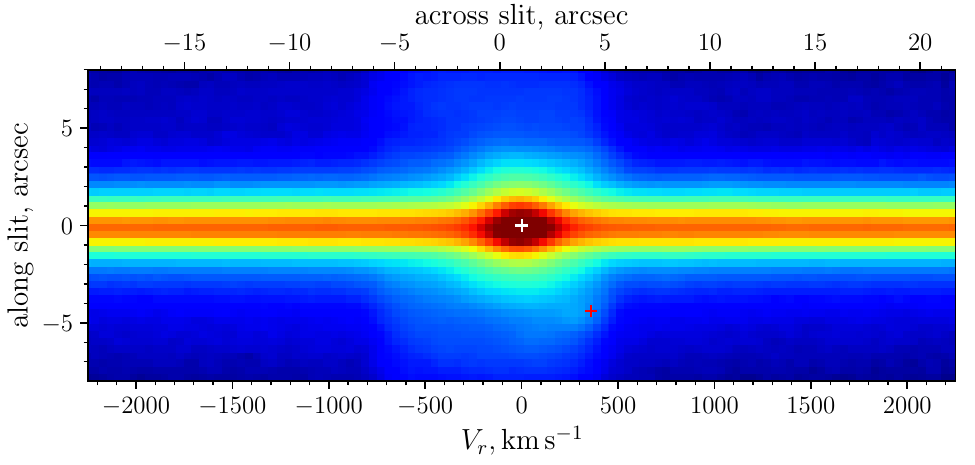}
      \caption{Wide-slit spectrum of DF Tau near H$\alpha$. The slit position is shown in Fig.~\ref{fig:vicinity}. The white cross marks the stellar position, and the red cross indicates the predicted position of the western blob at zero velocity. It can be seen that the actual position is blueshifted relative to the red cross.}
         \label{fig:blob-pvd}
   \end{figure}

We also found two H$\alpha$-emitting nebulae at a distance $d_{\text b}\approx 5\farcs 4$ from the star, located west and south of DF~Tau -- see the insert in the upper-left corner of the right panel of Fig~\ref{fig:vicinity}. The western blob (PA$\approx 265\degr$) is also visible in our 10\arcsec-slit spectrum of DF~Tau -- see Table~\ref{tab:log-TDS} and Fig~\ref{fig:blob-pvd}. We therefore conclude that at least the western nebulosity is a real structure, not an artifact of image processing. We know the position angle of the slit, the TDS image scale of $0\farcs 366$~px$^{-1},$ the dispersion of 39~\kms~px$^{-1}$ and the position of the star relative to the slit center, which we determined under the assumption that $V_{\text r}$ of the H$\alpha$ line in the stellar spectrum is zero. From this data, we estimated the velocity of the western blob as $V_{\text r}^{\text b}\sim -100$\,\kms. Assuming that the tangential velocity of the blob $V_{\text t}^{\text b}$ is approximately equal to its radial velocity $V_{\text r}^{\text b}$, the dynamical time of the blob is $t_{\text dyn}^{\text b}\sim d_{\text b}/V_{\text t}^{\text b}\sim 30$~yr.

This allows us to assume that the blob (or blobs?) could have been produced by the Year 2000-flare (Sect.~\ref{subsec:ring}). The PA of the western blob is noticeably different from that of the (micro-)jet. We note in this regard that the micro-jet was discovered from spectra observed before the year 2000, and furthermore, the blueshifted component of the [\ion{O}{i}]~6300~\AA{} line, with a velocity of $\approx-100$~\kms, is present in the spectrum of DF~Tau observed on December 6, 1984 \citep{Edwards-1987}, long before the Year 2000-flare. One can therefore assume that the blob is associated with a weakly collimated ejection, resembling something like a bubble in XZ~Tau \citep{Krist-XZTau-1997}.

\section{Concluding remarks }
 \label{sec:conclusion}

The H$\alpha$ images we obtained allowed us to detect a number of unusual emission nebulae in the vicinity of DF~Tau. We discovered a ring-shaped nebula around DF~Tau, but the binary system is not located at its center. We also identified a jet and a counter-jet emanating from the binary system; however, they are not antiparallel and move with different velocities. Consequently, it remains unclear whether this behavior arises from the  peculiarities of collimated flow formation from one or both components of DF~Tau. Additionally, it has become apparent that the HH objects of the HH~1266 flow are not arranged along a single smooth curve for some unknown reason.

In this paper, we have deliberately limited ourselves to a semi-qualitative interpretation of our observations at best. Deeper images and higher-quality spectra in various regions of the HH~1266 flow are needed for reliable conclusions. Additionally, information on the proper motion of the discovered nebulae and DF~Tau itself is required. One thing is certain: the binary is a source of both collimated and weakly collimated gas outflows with an unusual morphology and therefore deserves further study.

\begin{acknowledgements}
We thank the staff of the CMO SAI MSU for their assistance with the observations, Drs. A.~V.~Moiseev and K.~A.~Postnov for useful discussions, the referee for his helpful comments, and Prof. Bo~Reipurth for including the discovered HH-flow in the general catalog of objects of this type and assigning it the number HH~1266. This research has made use of the SIMBAD database (CDS, Strasbourg, France) and Astrophysics Data System (NASA, USA). The work of AVD (data reduction, interpretation) and IASh (spectroscopic observations) was conducted under the financial support from the Russian Science Foundation (grant 23-12-00092). Scientific equipment used in this study was bought partially through the M.~V.~Lomonosov Moscow State University Program of Development.
\end{acknowledgements}

\bibliographystyle{aa}
\bibliography{dftau.bib}

\end{document}